\documentclass[twocolumn, aps, groupedaddress, superscriptaddress, nofootinbib, showkeys, showpacs, notitlepage]{revtex4-1}

\usepackage{amsmath,amssymb}  % need for subequations
\usepackage{amsfonts}
\usepackage{graphicx,appendix}   % need for figures
\usepackage{hyperref}   % use for hypertext links, including those to external documents and UrLs
\usepackage{color}
\begin{document}

\title{Drive-specific adaptation in disordered mechanical networks of bistable springs} 
\author{Hridesh Kedia}
\affiliation{Physics of Living Systems Group, Department of Physics, Massachusetts Institute of Technology, Cambridge, MA 02139}

\author{Deng Pan}
\affiliation{Physics of Living Systems Group, Department of Physics, Massachusetts Institute of Technology, Cambridge, MA 02139}

\author{Jean-Jacques Slotine}
\affiliation{Nonlinear Systems Laboratory, Massachusetts Institute of Technology, Cambridge, MA 02139}

\author{Jeremy L. England}
\email{j@englandlab.com}
\altaffiliation{Senior Director in AI/ML, GlaxoSmithKline, 200 Cambridgepark Dr, Cambridge, MA 02140}
\affiliation{Physics of Living Systems Group, Department of Physics, Massachusetts Institute of Technology, Cambridge, MA 02139}

\begin{abstract}
Systems with many stable configurations abound in nature, both in living and inanimate matter.  Their inherent nonlinearity and sensitivity to small perturbations make them challenging to study, particularly in the presence of external driving, which can alter the relative stability of different attractors.  Under such circumstances, one may ask whether any clear relationship holds between the specific pattern of external driving and the particular attractor states selected by a driven multistable system. To gain insight into this question, we numerically study driven disordered mechanical networks of bistable springs which possess a vast number of stable configurations arising from the two stable rest lengths of each spring, thereby capturing the essential physical properties of a broad class of multistable systems. We find that the attractor states of driven disordered multistable mechanical networks are fine-tuned with respect to the pattern of external forcing to have low work absorption from it. Furthermore, we find that these drive-specific attractor states are even more stable than expected for a given level of work absorption. Our results suggest that the driven exploration of the vast configuration space of these systems is biased towards states with exceptional relationship to the driving environment, and could therefore be used to `discover' states with desired response properties in systems with a vast landscape of diverse configurations.
\end{abstract}

\maketitle

\section{Introduction}

Driven systems that are multistable appear in a diverse array of settings, whether in non-biological contexts such as nonequilibrium glasses \cite{tsamados_plasticity_2010, henderson_metastability_1996, berthier_non-equilibrium_2013}, driven disordered systems \cite{keim_generic_2011, keim_multiple_2013, paulsen_multiple_2014}, adaptable mechanical assemblies \cite{ harne_designing_2015}, artificial neural networks \cite{hertz_introduction_1991,stern_dynamics_2014,cheng_multistability_2006}, self-propelled bouncing drops \cite{bacot_multistable_2019}, lasers \cite{pisarchik_experimental_2003,pisarchik_control_2002}, DNA origami \cite{fern_energy_2016}, turbulent flows \cite{ravelet_multistability_2004}, and climate dynamics \cite{power_multiple_1993,rahmstorf_multiple_1995}, or in living matter such as in neural activity \cite{ braun_attractors_2010,marin_high_2013}, epigenetics \cite{ goldberg_epigenetics:_2007,laurent_multistability:_1999,wang_quantifying_2011}, protein-folding \cite{sittel_perspective:_2018}, and ecosystems \cite{knowlton_thresholds_1992,scheffer_catastrophic_2001}. 
% Is there another place to put this citation?, amongst a host of other systems \cite{feudel_complex_2008,pisarchik_control_2014}. 
Multistable systems, i.e. systems with coexistence of multiple dynamical attractors for a given set of system parameters \cite{feudel_complex_2008}, are very sensitive to initial conditions because their basins of attraction are usually complexly interwoven, with fractal boundaries \cite{feudel_complex_2008}. Their basins of attraction can undergo significant changes in the presence of a time-varying external drive, making it possible for the system's dynamical behavior to be shaped by its driving history. In particular, the presence of multiple stable states with different response properties to the external drive makes it possible for them to exhibit the emergent properties of memory \cite{hopfield_neural_1982,canavier_nonlinear_1993,bacot_multistable_2019,ravelet_multistability_2004,keim_generic_2011,keim_memory_2019} and adaptation \cite{kashiwagi_adaptive_2006,harne_designing_2015, bieling_force_2016, majumdar_mechanical_2018, pashine_directed_2019}.
 
 Past work on damped nonequilibrium dynamics \cite{england_dissipative_2015, perunov_statistical_2016} has pointed to the possibility that some dynamical attractors may be stabilized through dissipative adaptation---that is, as a result of sustaining exceptionally high work absorption during the system's history before reaching the attractor state. In systems with a diverse set of stable states, subjected to a pattern of external driving, a natural question is therefore whether the attractor states bear a signature of the specific pattern of driving. In this work, we study one possible mechanism for such an effect by simulating driven multistable mechanical spring networks.
 
Disordered mechanical networks subject to weak thermal noise are an especially attractive setting in which to study multistable driven dynamics because they capture the essential physical properties of a broad class of nonequilibrium systems of interest: their internal forces arise from a high-dimensional, rugged potential energy landscape possessing a vast range of different saddles and degenerate local minima, and their motion is dominated by the influence of drive energy that is constantly being absorbed as work and dissipated through drag; thus, such networks may offer insight into the physics of active matter, as well as the nonequilibrium self-assembly of materials with novel properties. Mechanical networks of bistable springs have previously been used as a model system for studying the glass transition \cite{yan_why_2013}, and for designing adaptable mechanical systems \cite{harne_designing_2015}.

The properties of a single, bistable one-dimensional oscillator are already quite illustrative of their capacity for exhibiting adaptive nonequilibrium behavior.  The spring potential energy has a double-well shape (see Fig.~1(a)), with the two local minima of differing curvatures corresponding to different rest lengths of the spring. In the weak noise limit, low-amplitude oscillatory driving with damping will yield two attractive periodic trajectories in the vicinity of each local minimum in the energy landscape.  At higher drive amplitude, however, it becomes possible for one trajectory to become unstable as a result of resonance, even as the other, less resonant one remains stable and becomes a global attractor. The history of work absorption through resonance therefore forms the basis for selection of a non-resonant state at long times.

\begin{figure*}[!htb]
\centering
\includegraphics[width=2\columnwidth]{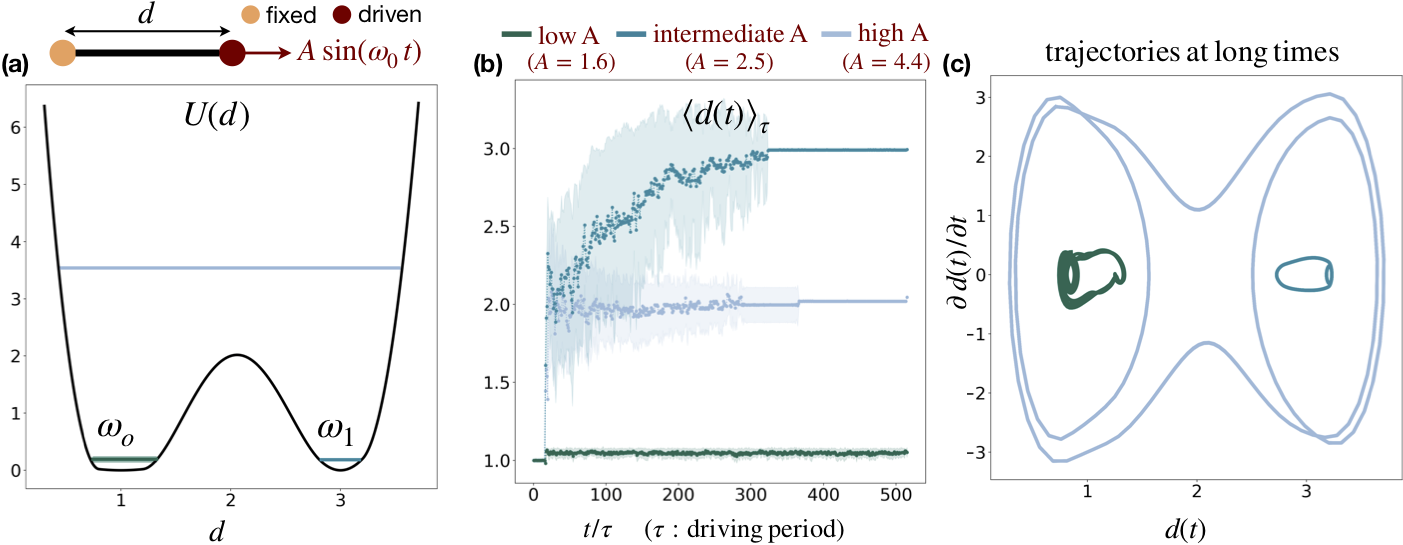}
\caption{(Color) One-dimensional bistable spring sinusoidally forced at low (green), intermediate (blue) and high (light blue) forcing amplitudes. (a) Bistable spring potential $U(d)$, with energy levels of final states for different forcing amplitudes. (b) Trajectory of average spring length, stuck near its initial short rest length for low amplitudes (green), stuck at the average of long and short rest lengths for high amplitudes (light blue), transitioning to the low work absorbing state of long rest length for intermediate amplitudes (blue). (c) Phase portrait of the system trajectory at long times. The averages are taken over $12$ runs starting from the short rest length, with different noise realizations.}
\label{bistable_1d}
\end{figure*}

This mechanism of annihilating an undesirable attractor via a harmonic perturbation at the frequency of its relaxation oscillations  \cite{pisarchik_annihilation_2000,pisarchik_controlling_2001, pisarchik_experimental_2003,pisarchik_control_2014} has already been developed in specialized applications for systems with relatively few, known metastable states \cite{pisarchik_controlling_2001,pisarchik_annihilation_2000, goswami_control_2009,pisarchik_experimental_2003,goswami_controlled_2007}. Here, we consider whether such resonant destabilization can act as a selection principle during the driven search of a high-dimensional space of possible metastable states for a disordered mechanical material.  Using numerical simulations, we study ensembles of disordered mechanical networks of bistable springs subjected to a time-varying external force and weak damping in the low-noise, strongly driven limit, where the topology of the disordered network is frozen. The quenched disorder of the network connectivity gives rise to a diversity of response properties for the different local energy minima that can act as metastable states of the system.  We reason that if each metastable state has its own distinct fingerprint of response properties, and if resonant response to oscillatory driving can lead to a high amplitude motion that triggers transition to a new state, then perhaps metastable states that have response properties well-suited to absorbing energy from the drive will be unlikely to be stable in the long run.

We demonstrate for several qualitatively different driving protocols that there is a range of forcing amplitudes for which the network becomes trapped in configurations whose response properties are adaptively fine-tuned to have atypically low work absorption from the drive. Once trapped in these configurations, the network is unlikely to absorb sufficient energy to exit the associated energy well. The maximum average work absorption rate of trajectories starting from such configurations which are fine-tuned to their drive is atypically lower than that of trajectories starting from the same configurations but paired with randomly chosen drives (see Fig.~\ref{atypicality_wdot}). We further demonstrate that the atypically low work absorption of the trapped configurations is a direct consequence of their response properties being fine-tuned to features of the time-varying drive, such as drive frequency and drive direction (see Fig.s~\ref{mode_analysis},\ref{changing_freq},\ref{changing_dirn}), and to the driving protocol (see Fig.s~\ref{pos_drive_intro},\ref{switch_force_pos}).

Moreover, we find that stable attractors are typically characterized by highly-correlated, low-dimensional motion of the particle collective, which enables greater stability than expected for the amount of energy being absorbed. Our results point to a general mechanism by which damped, driven many-particle systems with diverse local energy minima may explore their configuration space in a way that is biased towards states with atypically low energy absorption.

\begin{figure*}[!htb]
\centering
\includegraphics[width=2\columnwidth]{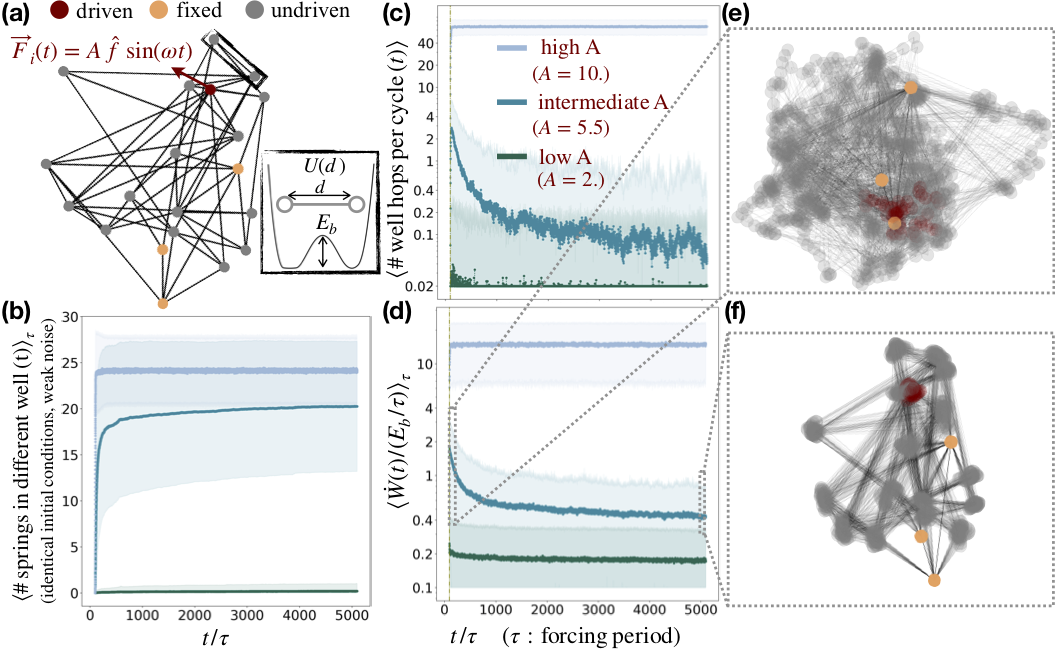}
\caption{(Color) Two-dimensional disordered mechanical networks of bistable springs, showing qualitatively different behavior for  different forcing amplitudes: low (green), intermediate (blue), and high (light blue). (a) An example of a disordered network of bistable springs. (b) Trajectories starting from identical initial conditions, subjected to the same forcing but different realizations of weak thermal noise, lead to different network configurations for intermediate (blue) and high forcing amplitudes (light blue). (c) Qualitatively different regimes of exploration of the landscape of network configurations. (d) Qualitatively different trajectories of average work absorption rate. The potential energy and the dissipation rate averaged over a drive cycle show similar trends as the work absorption rate as shown in the Appendix Fig.~\ref{force_drive_intro_supplement}. (e), (f) Overlay of snapshots from a sample trajectory, showing the network configuration at early times in (e), and at late times in (f) demonstrating the initial exploration of the configuration landscape and the eventual selection of a stable configuration for intermediate drive amplitude. Averages are taken over 1200 runs for each forcing amplitude: 240 networks each run 5 times with the same initial condition and different noise realizations.}
\label{force_drive_intro}
\end{figure*}

\section{A 1d driven bistable oscillator}
An illustrative setting showing the selection of a stable state with low work absorption is a driven one-dimensional bistable oscillator: two particles connected by a bistable spring with two rest lengths of equal energies but different spring stiffnesses, as shown in Fig.~\ref{bistable_1d}(a). While both rest lengths are equally stable in the absence of a drive, the presence of an external sinusoidal force at the resonant frequency of the shorter rest length makes the shorter configuration unstable because of its high work absorption due to resonance, while the longer rest length being far from resonance remains stable.

The equation of motion for the spring length $d$ is given by:
\begin{align}
&m\,\ddot{d}(t) = - U'(d) - \gamma\,\dot{d} + A\,\sin(\omega_0 t) + \xi(t) \label{1d_bistable} \\
&\langle \xi(t)\rangle = 0,\, \langle \xi(t) \xi(t')\rangle =2\gamma\,k_B T\,\delta(t-t')\; \nonumber
\end{align}
where the double-well spring potential $U(d)$ is shown in Fig.~\ref{bistable_1d}(a), with the natural frequencies of oscillation about the two rest lengths:$\omega_0=1\,,\,\omega_1=3.2$. The thermal noise is weak $(k_B T \ll E_b)$ and the dynamics are underdamped $(\gamma \ll m\omega_0)$.

For low forcing amplitudes, the spring length undergoes small amplitude oscillations about its initial short rest length $d=1$, while for high forcing amplitudes, the spring length keeps oscillating between short and long, i.e. bouncing back and forth between the high walls of the spring potential. However for intermediate forcing amplitudes, the high work absorption due to resonance in the shorter length state stretches the spring length beyond the potential energy barrier at $d=2$ and the spring length stabilizes in the vicinity of the longer rest length with low work absorption. 

\begin{figure}[!htb]
\centering
\includegraphics[width=\columnwidth]{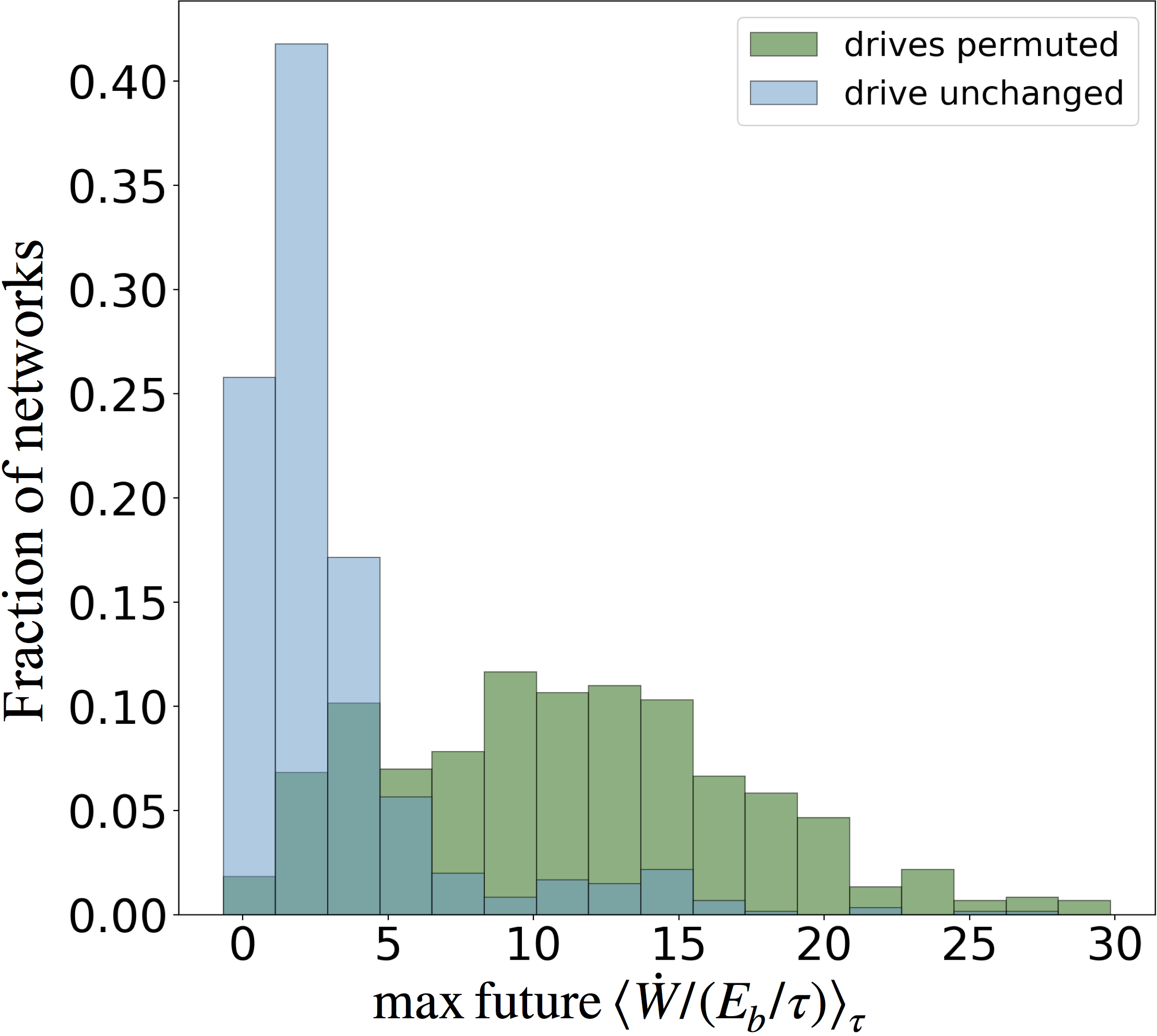}
\caption{\emph{Atypicality of work absorption rate at long times.} We run two sets of simulations of long duration (a few thousand drive cycles) starting from identical initial conditions, for an ensemble of $601$ disordered mechanical networks subjected to an ensemble of $601$ disinct drives of intermediate amplitude (A=5.5), with forcing frequencies uniformly spaced between $\omega=1$ and $\omega=7$, and forcing directions uniformly spaced between $0$ and $2\pi$. In one set of simulations, the drive is unchanged for the entire duration of the simulation, while in the other set the drives are permuted amongst themselves after half the duration of the simulation so that the system configuration is likely not fine-tuned to the new drive. The above plot shows the distribution of the maximum of the average work absorption rate for the later half of the system trajectory for both sets of simulations, demonstrating the significantly lower work absorption rate of the trajectories when the drive is unchanged compared to when the drives are permuted.}
\label{atypicality_wdot}
\end{figure}

In similar fashion to the behavior of this single spring model, instability of a stable state caused by a resonant perturbation was previously shown in the context of attractor annihilation \cite{pisarchik_controlling_2001,pisarchik_control_2014}, for low-dimensional dynamical systems with a few known attractors. In what follows, however, we demonstrate that even for many-body systems with many degenerate local energy minima in a high-dimensional energy landscape, orbits with low work absorption can be the global attractors as a result of the same physical mechanism.

\section{Driven disordered networks of bistable springs}
Disordered mechanical networks of bistable springs capture the essential physical properties of multistable systems whose metastable configurations span a diverse library of response properties. To gain insight into the relationship between the pattern of a time-varying external drive and the attractor states of such systems, we study ensembles of driven disordered mechanical networks of bistable springs using numerical simulations.

\begin{figure*}[!htb]
\centering
\includegraphics[width=2\columnwidth]{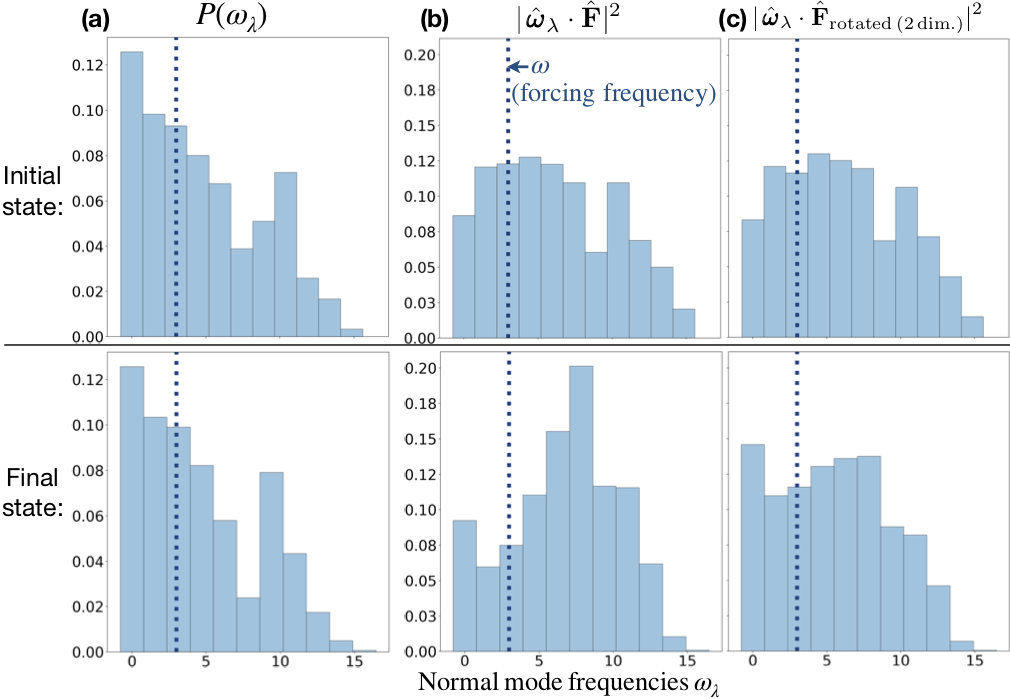}
\caption{Normal mode analysis of network configurations driven at intermediate amplitude (A=5.5), at the beginning and end of driving. (a) Density of normal modes, stays unchanged. (b) Coupling between normal modes and forcing, shows significant decrease for normal modes near resonance at end of driving. (c) Coupling between normal modes and rotated forcing (by $\pi/4$), does not show the same decrease for normal modes near resonance. Averages are taken over 1200 runs for $A=5.5$, as in Fig.~\ref{force_drive_intro}.}
\label{mode_analysis}
\end{figure*}

For the rest of this paper, we will consider two-dimensional disordered mechanical networks with $20$ particles (nodes), randomly connected by $50$ bistable springs (edges) each with a double-well spring potential (see Fig.~\ref{force_drive_intro}(a)). The disordered networks were generated from Erd{\"o}s-R{\'e}nyi graphs, although the qualitative conclusions of our study are true for small-world and scale-free networks as well as networks with different number of springs as shown in Appendix Fig.~\ref{diff_graphs}. The presence of many bistable springs and quenched disorder in the network imbues its high-dimensional potential energy landscape with many degenerate local energy minima. Each of these minima is a metastable configuration specified by the pattern of which edges in the network are close to their short or long equilibrium length.  Although not all of the $2^{50}$ such patterns are permitted by the network topology, the number of such metastable configurations is still vast, and each is characterized by a unique complement of principal directions---i.e. normal modes of vibration---and principal curvatures---i.e. natural frequencies of vibration---giving it a distinct fingerprint of response properties. Starting from a random metastable configuration, the network is subjected to a time-varying sinusoidal forcing, the energy absorbed from which is the primary driver of the exploration of its rugged potential energy landscape since the ambient thermal noise is weak. The work absorbed from the external force differs significantly across the different metastable states owing to their distinct response properties. Since transitions between metastable states must be activated by the work absorbed from the external force, we hypothesize that those with high work absorption are more likely to become unstable with driving.

The equation of motion for the $i^\mathrm{th}$ particle in the network is:
\begin{align}
m\,\ddot{\vec{x}}_i(t) &=-\sum_{j=1}^N \mathcal{A}_{ij}\vec{\nabla}_iU(\vert \vec{x}_i -\vec{x}_j \vert) -\gamma\,\dot{\vec{x}}_i + \vec{F}_i(t) \nonumber \\ 
&\qquad + \vec{\xi}_i(t) \label{2d_bistable_net}
\end{align}
where $N$ is the number of particles (nodes) in the mechanical network, and the adjacency matrix $\mathcal{A}_{ij}$ (encoding the topology of the disordered network), is independent of time. The external force $\vec{F}_i(t)=A\,\hat{f}\,\sin(\omega t)$ is applied to the most connected particle in the network, $\hat{f}$ denoting the direction of the forcing vector in the two-dimensional plane, while the three particles farthest from the driven particle (distance being the number of edges in the shortest path connecting two particles) are held fixed to constrain the global translations and rotations of the network. The thermal noise $\vec{\xi}_i(t)$ is weak i.e. $k_B T \ll E_b$, the dynamics are underdamped i.e. $\gamma \ll m\omega$, and are simulated using the Verlet integration scheme \cite{gronbech-jensen_simple_2013}. 
Analogous to the case of the single 1d bistable spring, for low forcing amplitudes, the system remains trapped in its initial configuration, while for high forcing amplitudes, the system keeps changing its configuration, exploring new metastable states in its vast energy landscape without ever attaining a stable configuration. However, for a range of intermediate forcing amplitudes, the system eventually becomes trapped in metastable configurations whose response properties are fine-tuned to have atypically low work absorption from the external forcing, preventing the system from ever acquiring sufficient input of work energy to exit the low-potential-energy absorbing state. %EXPLAIN
The atypically low work absorption of the configurations at long times indicates a specialized matching between the external forcing and the local response properties of the network state.

The driven exploration of metastable states over time is illustrated by the number of barrier crossings per drive cycle in Fig.~\ref{force_drive_intro}(c), where a barrier is crossed each time the length of a spring $d_{ij}$ changes past the energy barrier at $d_{ij}=2$ (see Fig.~\ref{force_drive_intro}(a)). The number of barrier crossings (well hops) per drive cycle is vanishing for low forcing amplitudes, and is always non-vanishing for high forcing amplitudes. However for intermediate forcing amplitudes, it is non-vanishing at early times and is vanishing at long times, indicating that a stable configuration, that is, a configuration in which none of the springs is changing length beyond the energy barrier at $d_{ij}=2$, is attained after an initial period of exploration. This account is confirmed by tracking the time-evolution of the potential energy (see Appendix Fig.~\ref{force_drive_intro_supplement}), which begins close to the global minimum value of zero, and rises transiently before becoming comparably low again at long times for intermediate forcing amplitudes, as the system becomes trapped in a stable configuration. The defining feature of these stable configurations attained at long times is their low rate of work absorption (see Fig.~\ref{force_drive_intro}(d)) and energy dissipation (see Appendix Fig.~\ref{force_drive_intro_supplement}).

The behavior of the system within the domain of a stable state discovered through driving is that of a periodic dynamical attractor. Because of the unexplorable vastness of the space of possible spring length patterns, we find that the same initial condition of the same random network experiencing the same external driving can settle into a diversity of different attractors depending on the realization of the weak thermal noise, as illustrated in Fig.~\ref{force_drive_intro}(b). However, although these stable configurations found at long times differ in their spring lengths, nearly all of them have in common low work absorption from the chosen external drive, as illustrated in Fig.~\ref{force_drive_intro}(d). 

The atypicality of the low work absorption rate within the stable states discovered through driving is demonstrated in Fig.~\ref{atypicality_wdot}, by comparing (i) trajectories in which the pairing between an ensemble of drives and networks is unchanged for the entire simulation, with (ii) trajectories that are identical to the previous set for their first half, but paired with the drives randomly permuted for their second half. The stable configurations attained at the end of the first half of the trajectories are fine-tuned to the drives experienced until then, leading to atypically low values for their maximum future work absorption rate when the drives are unchanged as compared to when the drives are permuted. 

\begin{figure*}[!htb]
\includegraphics[width=2\columnwidth]{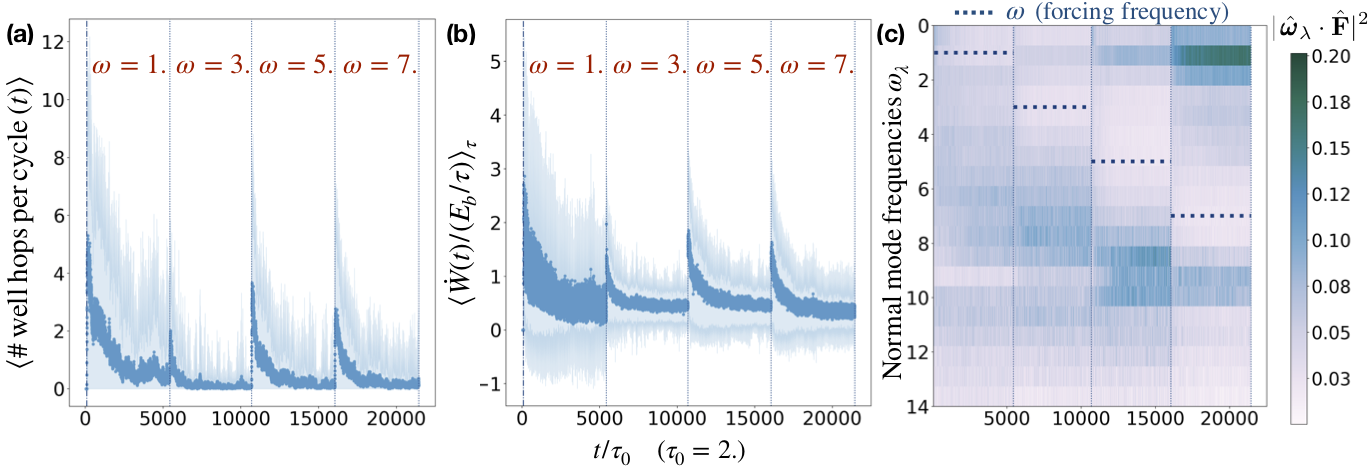}
\caption{(Color) Changing forcing frequency leads to a new configuration fine-tuned to have low work absorption at new forcing frequency. (a) A switch in forcing frequency causes a spike in the number of barrier crossings per cycle. (b) The reason for more barrier crossings is the spike in the work absorption rate. (c) The coupling of normal modes with the forcing, becomes atypically weak for modes near the forcing frequency, thereby reducing the work absorption rate, as in Eq.~\ref{wdot_lr}. Averages are taken over 120 networks.}
\label{changing_freq}
\end{figure*}

\emph{Mechanical linear response properties.} To further understand the low work absorption of the configurations at long times, we study their mechanical linear response properties using a normal mode analysis. The work absorption being low at long times, the external drive causes only small changes in the spring lengths, permitting a linear approximation of the forces generated by the bistable springs, to give a linear equation of motion:
\begin{align}
m\,\ddot{\vec{x}}_i(t) &=-\sum_{j=1}^N \overset{\text{\tiny$\leftrightarrow$}}{D}_{ij}\cdot \vec{x}_j -\gamma\,\dot{\vec{x}}_i + \vec{F}_i(t) + \vec{\xi}_i(t) \,, \label{2d_bistable_net_linear}
\end{align}
where the dynamical matrix $\overset{\text{\tiny$\leftrightarrow$}}{D}$ is:
\begin{equation}
D_{ij}^{ab}=\begin{cases}
-\mathcal{A}_{ij}\,U''(\vert \vec{X}_{ij} \vert)\,\hat{X}_{ij}^{a}\,\hat{X}_{ij}^{b} \;,& i\neq j \\
\sum_k \mathcal{A}_{ik}\, U''(\vert \vec{X}_{ik}\vert)\,\hat{X}_{ik}^{a}\,\hat{X}_{ik}^{b} \;,& i= j 
\end{cases} \label{linear_Dmat}
\end{equation}
where $\vec{X}_{ij}=\vec{x}_i-\vec{x}_j$ and $\hat{X}_{ij} = \vec{X}_{ij}/\vert \vec{X}_{ij} \vert$. The distinct fingerprint of response properties of a metastable configuration of the system is governed by the local shape of the associated high-dimensional potential energy well, which in the linear approximation is characterized by the normal modes of vibration $\hat{\pmb{\omega}}_\lambda$ (principal directions) and the natural frequencies of vibration $(\omega_\lambda)$ (principal curvatures), i.e. the eigenvectors and eigenvalues of the dynamical matrix $\overset{\text{\tiny$\leftrightarrow$}}{D}$. The linear approximation allows an exact calculation of the average work absorption rate over a drive cycle:
\begin{align}
    \langle \mathbf{F}(t)\cdot \mathbf{V}(t)\rangle_\tau &= \frac{\gamma\,A^2\,\omega^2}{2}\sum_{\lambda=1}^{2N} \frac{(\hat{\mathbf{F}}\cdot\hat{\pmb\omega}_\lambda)^2}{m^2(\omega_\lambda^2-\omega^2)^2 + \gamma^2\omega^2} \label{wdot_lr}
\end{align}
where $\mathbf{F}(t),\,\hat{\mathbf{F}},\,\mathbf{V}(t)$ are $2N$-dimensional vectors consisting of the $x,y$ components of the forces and velocities for each of the $N$ particles. The contribution of each normal mode $\hat{\pmb\omega}_\lambda$ to the work absorption rate in Eq.~(\ref{wdot_lr}), is greater the closer it is to resonance, i.e. the smaller $(\omega_\lambda^2-\omega^2)^2$ is, and is controlled by how strongly the forcing couples to it, i.e. $(\hat{\mathbf{F}}\cdot\hat{\pmb\omega}_\lambda)^2$. The normal modes that are close to resonance typically contribute the most to the work absorption rate. It should be emphasized however, that while this linear approximation is informative with respect to the energy flow of the stable attractors in this system, the transient search of configuration space that precedes the formation of such attractors is highly non-linear, and has no analog in the linear response regime.  Indeed, it is the presence of non-linearity that enables the system to tune the local linear response properties to match the drive.

A normal mode analysis of the intial network configurations and the network configurations after a long period of driving, illustrated in Fig.~\ref{mode_analysis}, shows that while the density of normal modes close to resonance shows no significant change, the coupling between the external forcing and the normal modes close to resonance reduces significantly. In other words, at long times the external forcing is unable to excite the normal modes close to resonance as strongly as in the initial configuration, leading to lower work absorption from the drive. That this signature of the forcing frequency disappears for a rotated version of the forcing (see Fig.~\ref{mode_analysis}(c)) demonstrates that the network configuration at long times is fine-tuned not only to the forcing frequency but also to the forcing direction.

\subsection{Changing drive complexity}
\emph{Changing forcing frequency, direction.} The signatures of features of the external forcing are a result of the network being fine-tuned to have atypically low work-absorption from it. %EXPLAIN
A change in these features of the forcing should lead to higher work absorption and a renewed exploration of configuration space until a new metastable configuration with low work-absorption is attained. Consistent with this expectation, a sudden change in the forcing frequency is observed to lead to an increase in the work absorption rate accompanied by a renewed exploration of configuration space, and the eventual selection of a state with low work-absorption, as demonstrated in Fig.~\ref{changing_freq}. The same phenomenon is observed for sudden changes in forcing direction, as shown in Appendix Fig.~\ref{switching_dirn}.

\begin{figure*}[!htb]
\includegraphics[width=2\columnwidth]{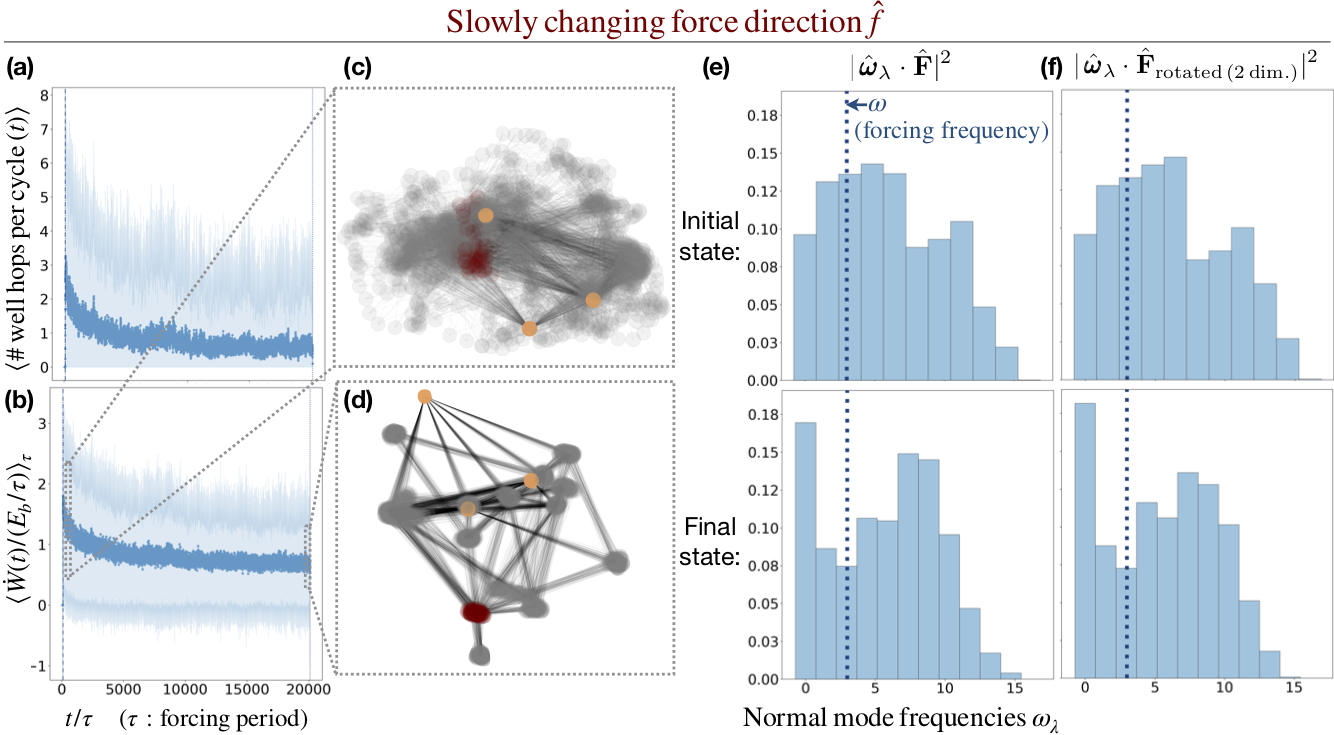}
\caption{(Color) Changing the direction of forcing every $5$ cycles by a Gaussian distributed angle $\mu=0,\sigma=\pi/10$. (a) Changes to spring configurations reduce over time as the network attains a stable configuration. (b) Average work absorption rate is atypically low in the stable configuration. (c),(d) Overlay of snapshots of a network configuration at early times (c) and at late times (d) demonstrating the initial exploration of the configuration landscape and the eventual selection of a stable state. The coupling between the normal modes near resonance and the forcing as well as a rotated forcing (by $\pi/4$) is significantly reduced at the end of driving, reducing the work absorption rate for forcing across a range of directions. Averages are taken over 180 networks.}
\label{changing_dirn}
\end{figure*}

Thus far, we have only considered an external forcing whose frequency and direction were both constant for long periods of time, allowing the system to attain a configuration fine-tuned to these features. However, even when the direction of the $2$-dimensional forcing vector $\hat{f}$ is constantly subject to small random changes, remarkably, a metastable configuration that is fine-tuned to have low work-absorption from the constantly changing forcing is typically discovered at long times. In this case, the forcing vector as well as its rotated version show atypically weak coupling to the normal modes near resonance as shown in Fig.~\ref{changing_dirn}, suggesting that these configurations have low work absorption for any direction of forcing on the driven particle.

\emph{A more challenging driving protocol---controlling position.}
For each of the time-dependent external drives described above, the driven exploration of the vast landscape of metastable configurations successfully selects at long times, a configuration whose response properties are fine-tuned to have atypically low work absorption from the external drive. %EXPLAIN 
A common feature of the external drives considered so far was that the force exerted on the mechanical network was dependent on time, but independent of its configuration.

\begin{figure*}[!htb]
\centering
\includegraphics[width=2\columnwidth]{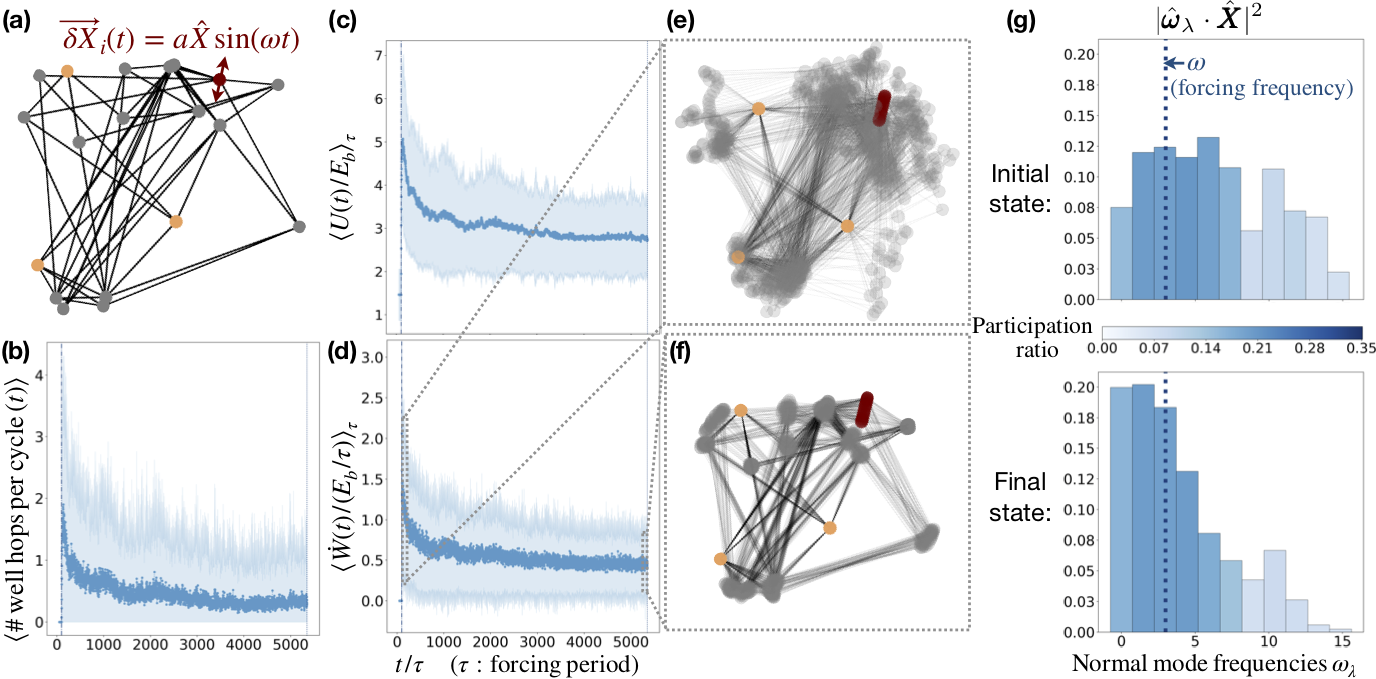}
\caption{(Color) Sinusoidal changes in the position of the driven particle, instead of sinusoidal external force. (a) The most connected particle of the network is displaced sinusoidally along a fixed direction at a fixed frequency, at an intermediate amplitude (a=0.2). The number of springs changing their configuration in (b), the potential energy in (c), and the work absorption rate in (d) reduce over time as the network attains a stable configuration. A consistent feature of the stable configuration is its atypically low work absorption. Overlays of snapshots of the network at early times in (e), and at late times in (f), demonstrate the initial exploration of the landscape and the eventual selection of a stable state. The coupling of the normal modes with the driving displacement vector increases significantly for the soft (low frequency) normal modes thereby reducing the force required to displace the driven particle. The soft modes are also those involving collective motions of the network as shown by their higher participation ratio. Averages are taken over 120 networks.}
\label{pos_drive_intro}
\end{figure*}

We now consider a more challenging driving protocol in which the position of the driven particle undergoes a sinusoidal displacement of an intermediate amplitude, along a fixed direction at a fixed frequency, as illustrated in Fig.~\ref{pos_drive_intro}. In this driving protocol, the force required to ensure the sinusoidal displacement of the driven particle depends not only on time but also on the configuration of the mechanical network, making the search for a configuration with low work absorption more challenging. Our numerical simulations show that even in this case, the metastable configurations attained at long times are fine-tuned to have atypically low work absorption from the drive. %EXPLAIN 

The normal mode analysis of the metastable configurations, illustrated in Fig.~\ref{pos_drive_intro}(d) reveals that the coupling of the drive displacement to the low frequency (soft) collective modes of the mechanical network is greatly enhanced at long times, thereby reducing the force needed for the sinusoidal displacements of the driven particle and the work absorbed from the drive. The enhanced coupling of the drive to soft collective modes is qualitatively distinct from the reduced coupling of the drive to modes near resonance for sinusoidal forcing of the driven particle. This qualitative distinction between the configurations at long times of the displacement-driven mechanical networks and the force-driven mechanical networks is a result of the system having adapted in distinct ways to experience low work absorption from these distinct ways of driving.

\begin{figure*}[!htb]
\centering
\includegraphics[width=2\columnwidth]{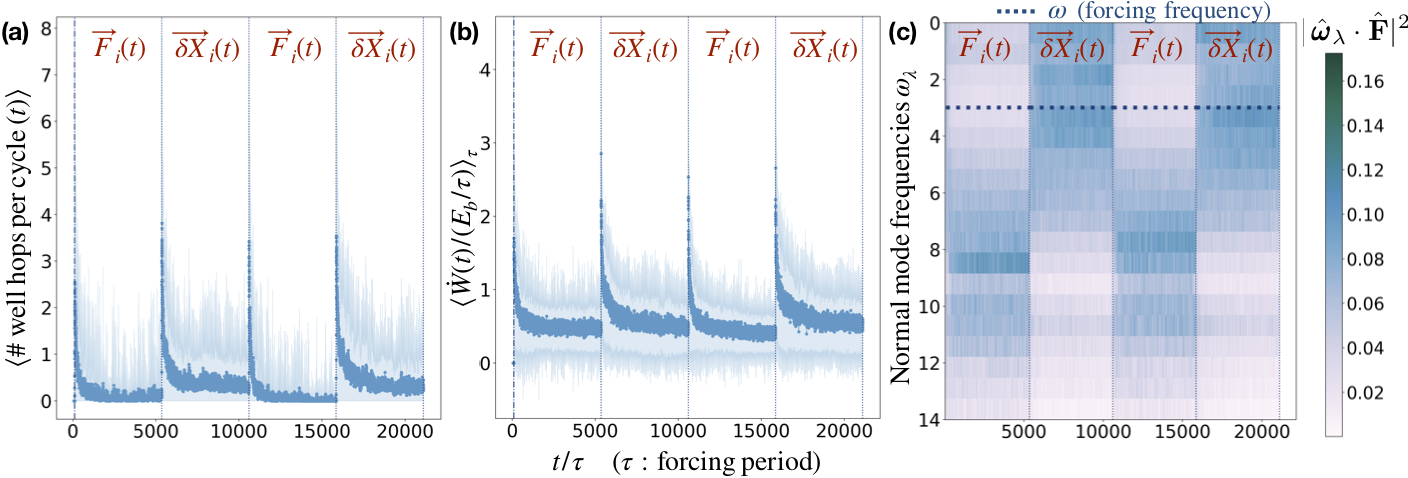}
\caption{(Color) Switching between driving protocols, sinusoidal forcing and sinusoidal displacement of driven particle, while keeping driving direction and frequency fixed. (a) A switch in driving protocol leads to a renewed exploration of the configuration landscape, eventually attaining a new stable configuration. (b) A switch in driving protocol leads to a spike in the work absorption rate powering the landscape exploration, eventually attaining a stable configuration with atypically low work absorption. (c) The coupling between normal modes and the driving vector changes each time the driving protocol changes, ensuring low work absorption of the stable configuration for the new driving protocol. Averages are taken over 120 networks.}
\label{switch_force_pos}
\end{figure*}

To verify the qualitatively distinct character of the force-driven and displacement-driven configurations at long times, the two driving protocols are applied for long periods of time, one after the other, without any change in the drive frequency or direction. Our simulations show that each time the drive protocol is switched, the work absorption rate increases, leading to a renewed exploration of configuration space and the eventual selection of a new configuration with low work absorption. This confirms the differing character of the low work-absorbing configurations for the force-driven and displacement-driven mechanical networks, and demonstrates the flexibility possible in the shaping of the response properties of a system with many metastable configurations by an external drive.

 For the variety of driving protocols described above, we observe the selection at long times of particular metastable attractor states from a vast high-dimensional landscape of possibilities.  The fine-tuned response properties of these attractors suggest that the driven exploration of configuration space is biased towards configurations with atypically low work absorption. %EXPLAIN
 This dissipation-reducing effect is loosely reminiscent of a well-known requirement, due to Prigogine, that linear-regime nonequilibrium steady-states must be local and global minima of entropy production \cite{prigogine_etude_1947}. However, the physical mechanism for the effect identified here is quite distinct: in disordered, damped many-particle systems such as the ones considered above, it is only as the result of \emph{non}linearity in the energy landscape that a barrier-jumping search of different local minima can ultimately lead to a decrease in dissipation.

\begin{figure*}[!htb]
\centering
\includegraphics[width=2\columnwidth]{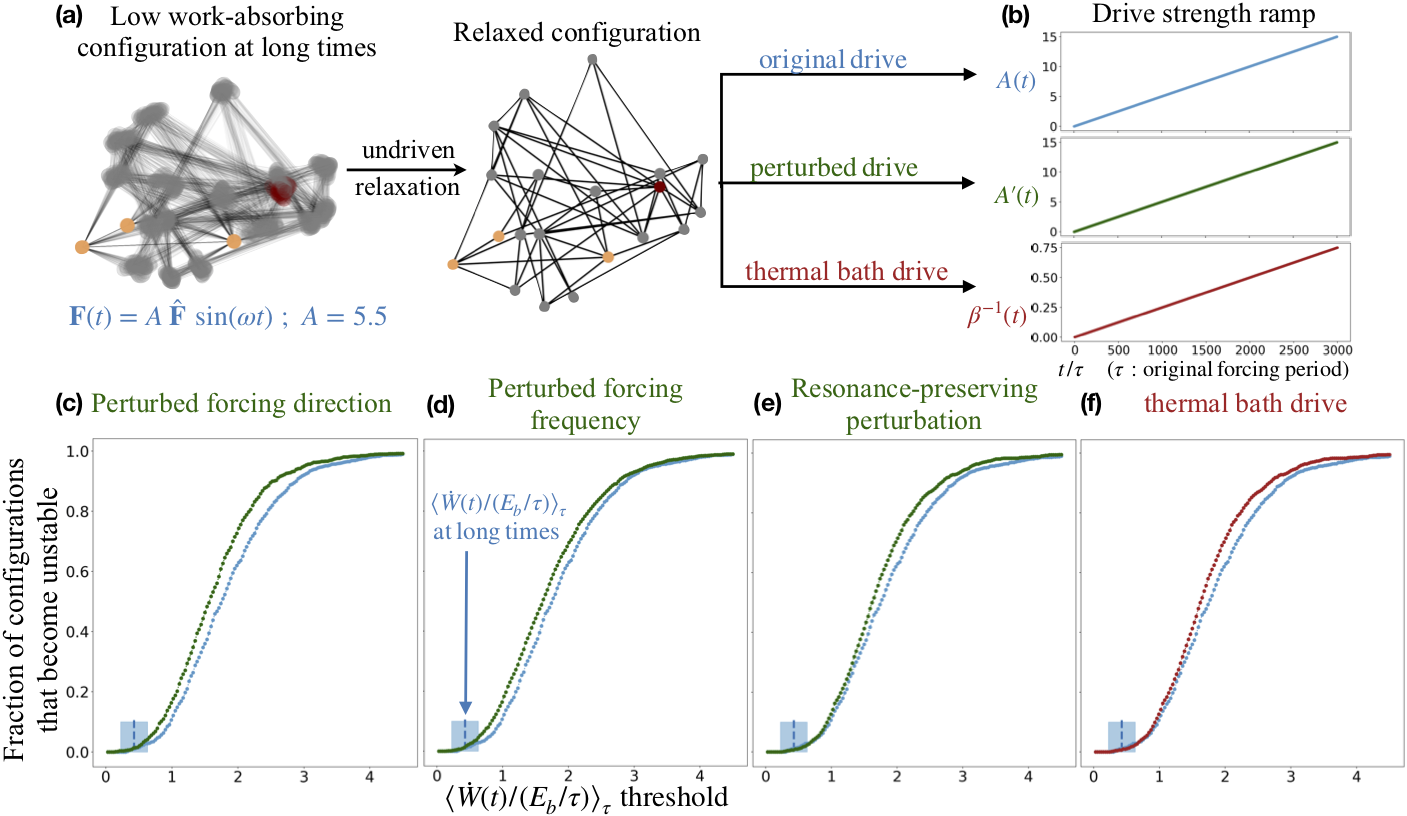}
\caption{(Color) Comparison of the stability of the driven motion in the configurations attained after a long period of driving, by applying different drives with increasing drive strength. (a) Configurations fine-tuned to have low work absorption from the drive, at the end of a long period of sinusoidal forcing at intermediate amplitude (shown in Fig.~\ref{force_drive_intro}), relax to the bottom of their potential energy well in the absence of a drive. (b) With the relaxed configuration as initial condition, different drives with increasing drive strength are applied. Among these drives are the drive with which the network was originally driven, perturbed versions of that drive, and a thermal bath drive where all particles in the network are driven by thermal forces from an external thermal bath. (c)-(f) The fraction of driven trajectories that become unstable before hitting a work absorption rate threshold is lower for the original drive than for perturbed drives or the thermal bath drive. (c) The perturbed forcing is along a direction $\pi/3$ rotated from the original forcing direction. (d) The perturbed forcing is at a frequency $\pm 33\%$ shifted from the original forcing frequency. (e) The perturbed forcing is obtained by rotating the projection of the original forcing direction in the space of two nearly degenerate normal modes by $\pi$, ensuring that the work absorption rate in Eq.~(\ref{wdot_lr}) remains unchanged. (f) The entire network is subjected to random thermal forces. Data from 1157 simulations.}
\label{enhanced_stability}
\end{figure*}

\subsection{Stability of the fine-tuned metastable configuration} 
We have demonstrated for a variety of driving protocols that, after a long period of driving, disordered mechanical networks of bistable springs are typically trapped in a metastable configuration---a potential energy well---which is fine-tuned to the external drive to have low work absorption from it. The truth is, however, that such attractors are not selected for low energy absorption, \emph{per se}, but rather for their inability to escape one particular region of phase space.  Though the motion with lower velocity and amplitude that results from lower work absorption clearly increases the likelihood of becoming confined, it is also conceivable that the shape of the dynamical trajectory might play a role as well.

The significance of this possibility becomes clearer once the system is viewed through the lens of contraction analysis \cite{lohmiller_contraction_1998}. Within a single potential energy well where the Newtonian potential is entirely concave up, there always exists a region for which every trajectory in that locale in phase space is converging towards a single trajectory which is particular to the external drive and the confining potential well. For periodic driving, this trajectory must be a closed periodic orbit, in other words, a one-dimensional object to which the coordinated motion of many particles has become confined. Since such motion is constrained to a far greater extent than thermalized random motion at the same energy, it is possible for a trajectory to be trapped by the shape of the surrounding energy landscape only because its high amplitude motions are well-matched to high energy barriers. By comparing the stability of the system across a range of work absorption rates, for different trajectories (corresponding to different external drives) within the same potential well, we therefore have the opportunity to demonstrate a separate way in which the same selection principle acts to finely-tune the attractor states.

We take as our starting point the stable low work absorbing configurations discovered by driving the system with a sinusoidal force of intermediate amplitude at a fixed frequency and along a fixed direction, illustrated in Fig.~\ref{force_drive_intro}. After allowing the system to relax to the bottom of its potential energy well in the absence of an external drive, multiple simulations with different drives are run starting from the relaxed initial condition. The drives include the original drive which discovered the potential well, a drive with a $33\%$ difference in forcing frequency, a drive whose forcing direction is rotated by $\pi/3$, a drive which is rotated in the space of nearly degenerate normal modes to have nearly the same linear response work absorption rate given in Eq.~\ref{wdot_lr}, and a drive which is an additional thermal bath. Each of the different drives corresponds to a different system trajectory in the potential well. The amplitude of each external drive is slowly increased from low to high, thereby increasing the work absorption rate and eventually making the trajectory become unstable and exit the potential energy well. We study the stability of the different trajectories at different levels of work absorption by analyzing the probability of exiting the potential well before the work absorption rate reaches a given value. We find that the fraction of simulations that become unstable before a given value of the work absorption rate is attained is lower for the original drive than for any of the other drives, as shown in Fig.~\ref{enhanced_stability}. This suggests that at the same level of work absorption, the system trajectory corresponding to the original drive is atypically stable.

\section{Discussion}

In this study, we have shown that a simulated, two-dimensional, multi-stable, mechanical network with disordered connectivity can exhibit a flexible adaptive capability, leading to emergent response properties that are finely-tuned with respect to the pattern of external forcing.  Over multiple realizations of connectivity disorder and thermal noise, we have demonstrated a reproducible tendency for such systems to settle into states that, though typically different in their microscopic details, have in common the macro-property of atypically low work absorption from the external drive.  This adaptive reduction in work absorption is responsive to changes in direction and frequency of the driving, as well as to shifts from applied sinusoidal forcing to sinusoidal positioning for the driven particle.  Moreover, we have found that the drive-specific stable states discovered by the driven dynamics in these systems are even more stable than would be expected given their level of work absorption, indicating that their response properties are also finely tuned in shape to reduce the activation of barrier-crossing motions.

There are a number of examples of many-body self-organization that see systems increase in their ability to absorb work from the external environment over time \cite{kondepudi_end-directed_2015,kachman_self-organized_2017}, but such ``energy-seeking" behavior is by no means the only example of emergent fine-tuning in nonequilibrium dynamics. Here, we have investigated a different, but equally important aspect of the physics of self-organization, namely: the emergence of fine-tuned structure that can remain stable in the presence of potentially disruptive energy supplied from a patterned environmental source.  Whether in the case of increasing or decreasing energy absorption, past studies have referred to such emergent fine-tuning behaviors as dissipative adaptation. The hallmark of this effect is that likely configurations for the system at long times have energy absorption properties that reflect a history of undergoing irreversible reconfiguration during elevated absorption and dissipation of work from the drive \cite{perunov_statistical_2016}.  The case examined here is the first to be indentified in a simple Newtonian equation of motion for a classical many-body systenm, and resultingly, the role of dissipation in the mechanism of fine-tuning is relatively straightforward and intuitive: the networks jump between local degenerate energy minima by absorbing work as potential energy during traversal of a barrier, and then dissipating it through drag as the system settles into a new minimum.  Irreversibility is made especially possible when work absorption is exceptionally high, since the system is unlikely to absorb enough work to go back the way it came once resonant absorption drops after traversal of a barrier.

It should be emphasized that the reduction in absorption and dissipation of work observed in this study is wholly distinct from the minimization of entropy production expected in a linear-response regime nonequilibrium steady-state.  The simulations reported here were carried out in the highly nonlinear-response regime, and are not constrained by Onsager relations to minimize entropy production in steady-state.  Moreover, the space of possible arrangements for the spring lengths of the network is so vast that a steady-state sampling of these metastable states cannot be achieved for such low temperatures; rather, the low dissipation states discovered by the dynamics behave like absorbing attractors.

Accelerated exit from specific subsets of states has been identified previously as a possible selection mechanism for stable attractors in instances with strong time-scale separation between driven fast degrees of freedom and reversibly-coupled slow ones \cite{chvykov_least-rattling_2018}. 
In such a ``least-rattling" picture, fluctuations from fast motion act as a spatially-varying diffusion coefficient for the slow variables, which tend to settle where diffusion is weakest.  In the system considered here, however, there is a single, simple Hamiltonian governing one kind of particle degree of freedom, and thus there is no explicit separation between fast and slow coordinates.  Moreover, due to the low temperature limit, the driven dynamics are nearly deterministic, and become completely so in the absence of thermal noise.  Thus, the results reported here demonstrate that selection of attractors based on the degree to which and on the way in which they absorb energy from a drive stands on more general footing than the original least-rattling scheme, and can govern the behavior of a broader class of many-body dynamics.

Several phenomena that operate by a similar physical mechanism have been previously identified. In resonant destruction of attractors of multi-stable lasers \cite{pisarchik_control_2002,pisarchik_experimental_2003}, and in spectral hole burning of optically stimulated materials \cite{volker_hole-burning_1989}, it has been established that the external driving can destroy states with high energy absorption. In the case of spectral hole burning, the molecules that absorb resonantly at the incident frequency undergo a phototransformation so that they absorb at a different frequency \cite{volker_hole-burning_1989}, while in multi-stable lasers the attractors with relaxation frequencies that match the driving frequency are annihilated. However, the dynamics of the systems in these past examples are constrained to be responsive only to the drive frequency and result in the destruction of high energy absorbing states, giving them limited flexibility in their adaptive response.

In the limit of zero temperature, the networks studied here act as deterministic, damped Hamiltonian systems.  As a result, there is a straightforward relationship between local properties of the potential energy landscape and the degree of contraction experienced by neighboring trajectories in phase space. In particular, it can be shown that the number of concave up (positive curvature) directions determined in the Hessian of the potential corresponds to the number of locally contracting directions, which means all local energy minima have a fully-contracting region of finite volume surrounding them~\cite{lohmiller_contraction_1998}.
Since the internal force on the system always points in the direction of decreasing potential energy, the dynamics generally have a tendency to seek out contracting regions, and this manifests in a tendency for the many-body motion of the system to become confined to a low-dimensional sub-volume of its configuration space.  Indeed, in the extreme limit where the system settles into one local energy minimum and cannot exit, the motion is restricted to a one-dimensional periodic attractor along a closed loop in configuration space, which for $N\gg 1$ particles corresponds to the emergence of highly coordinated, correlated motion involving many degrees of freedom.  

It is significant on its own that this level of coordinated motion emerges necessarily as a consequence of the fact that the system has found a fine-tuned state that can keep its energy low enough to avoid barrier jumping.  However, our perturbation studies have revealed that the contraction to low-dimensional dynamics may actually cooperatively assist in the further stabilization of the fine-tuned state.  Not only does the mechanical network reduce its overall energy absorption, it also deploys the energy it does absorb into motions that cannot disrupt its pattern of spring lengths.  Thus, when a drive to which the system did not adapt delivers the same amount of energy into the particles, they are more likely to move in ways that activate large-scale rearrangements and accompanying changes in response properties. It is notable that in both technological and biological contexts, there is a usual way energy is absorbed that drives functional motion, whereas improperly deployed energy is much more likely to cause random and larger-scale structural arrangements that are often referred to as damage -- for example, a living cell can power itself by digesting sugar, but cannot be be nourished by absorbing an equivalent quantity of energy in the form of heat or gamma radiation.  In this study, we are confronted with an example of an excited, many-body nonequilibrium state that exhibits a similarly specific relationship to energy input from its environment, and the origin of this specificity is simply that the ``preferred" energy source gave rise to the stable structure in the first place.

\section{Conclusion}
We have demonstrated for ensembles of disordered mechanical networks of bistable springs subjected to a variety of driving protocols, a fine-tuning between the drive pattern and the driven attractor states which ensures atypically low work-absorption from the drive. That the system dynamics `discover' such fine-tuned states in the vast landscape of configuration space suggests that the driven exploration of configuration space is biased towards states with atypically low work absorption. Furthermore, we have demonstrated that the driven motion of the network in these attractor states is atypically stable for the same level of work absorption.

It should be emphasized, in closing, that the physical mechanism we demonstrate here derives from highly generic ingredients: in theory, all that is required is a multi-stable arrangement of many interacting particles whose rugged potential energy landscape exhibits a diversity of response properties in different local minima. In such a system, patterned driving within a certain range of amplitudes may easily give rise to selection of attractor states that exhibit fine-tuned matching. Accordingly, our initial investigations here imply a wide-range of experimental settings where it may be fruitful in future to search for and characterize similar effects. Such fine-tuning might eventually be harnessed to `discover' structures and materials with desired response properties via the driven search of a large configuration space, with exciting possibilities for self-assembly, growth processes, and pattern formation.

\begin{acknowledgments}
The authors acknowledge valuable discussions with Pavel Chvykov, Jeremy Owen, Jacob Gold, Sumantra Sarkar, Gili Bisker, Jordan Horowitz, and Benjamin Machta. J.L.E. is funded by the Air Force Office of Scientific Research grant FA9550-17-1-0136 and by the James S. McDonnell Foundation Scholar Grant 220020476. H.K. is funded by the Air Force Office of Scientific Research grant FA9550-17-1-0136.
\end{acknowledgments}

\appendix*
\begin{figure*}[!htb]
\centering
\includegraphics[width=2\columnwidth]{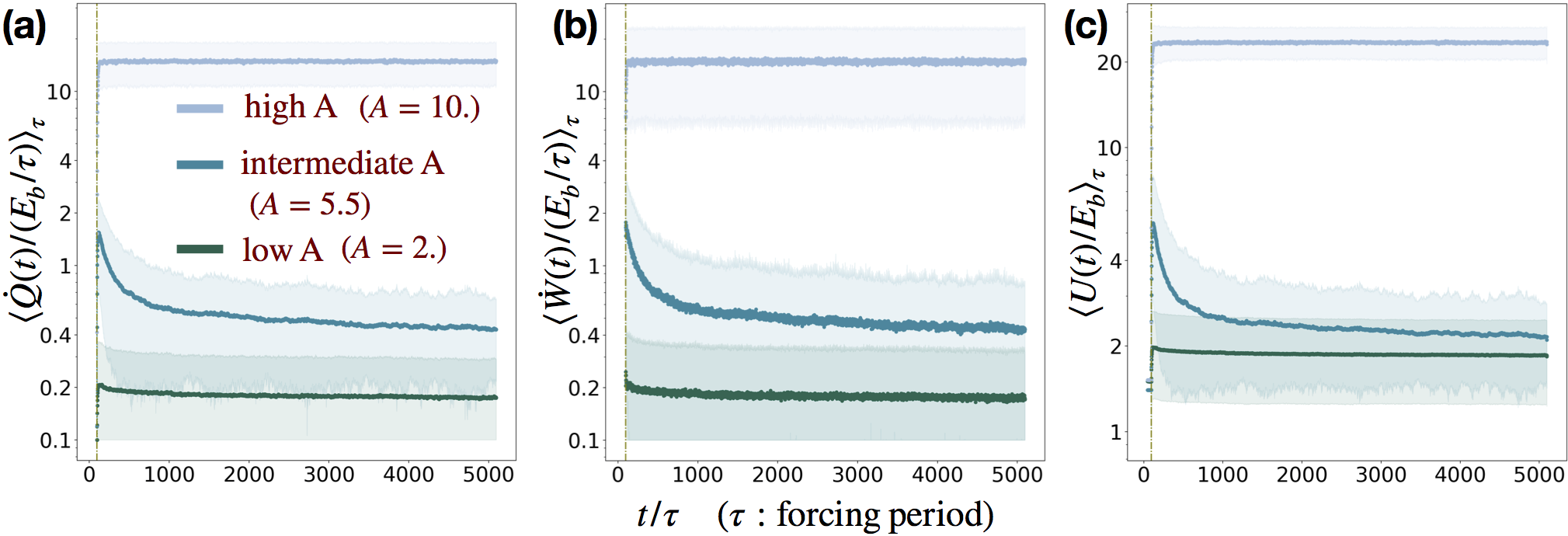}
\caption{Supplement to Fig.~2 of main text. Two-dimensional disordered mechanical networks of bistable springs, generated from Erd{\"o}s-R{\`e}nyi graphs with $20$ nodes and $50$ edges as in Fig.~2, showing qualitatively different behavior for  different forcing amplitudes: low (green), intermediate (blue), and high (light blue). 
(a) The dissipation rate averaged over a drive cycle, (b) the work absorption rate averaged over a drive cycle, and (c) the potential energy averaged over a drive cycle show the same trends, staying low for low amplitude drive (green), staying high for high amplitude driving (light blue), and for intermediate amplitude driving (blue), they are initially increasing as the system explores its vast configuration landscape, and decreases to a low value as a stable configuration with low work-absorption rate, low dissipation rate, and low potential energy is attained.}
\label{force_drive_intro_supplement}
\end{figure*}

\begin{figure*}[!htb]
\centering
\includegraphics[width=2\columnwidth]{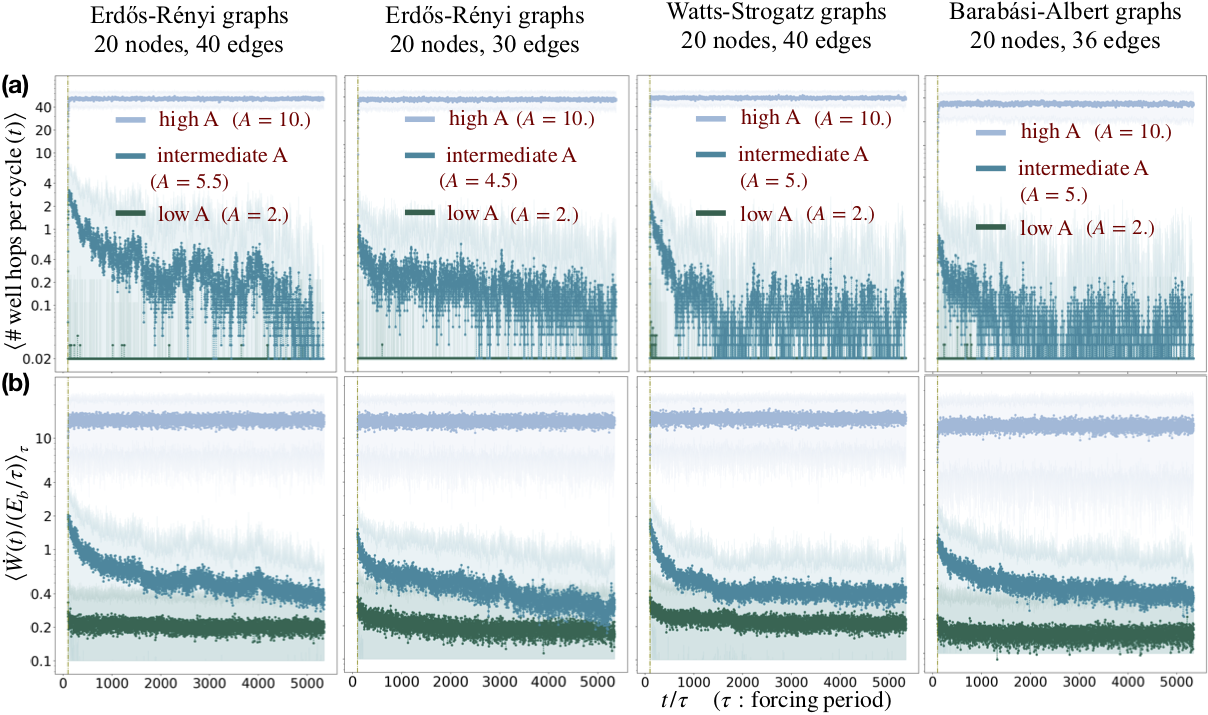}
\caption{Two-dimensional disordered mechanical networks of bistable springs, generated from different types of graphs, showing qualitatively different behavior for  different forcing amplitudes: low (green), intermediate (blue), and high (light blue). Similar behavior as in Fig.~2 main text, is observed as a function of forcing amplitude for a wide variety of networks, suggesting the wide applicability of the observations. (a) Qualitatively different regimes of exploration of landscape of network configurations. (b) Qualitatively different trajectories of average work absorption rate. Averages are taken over 100 networks for each forcing amplitude, and each type of graph.}
\label{diff_graphs}
\end{figure*}

\begin{figure*}[!htb]
\centering
\includegraphics[width=2\columnwidth]{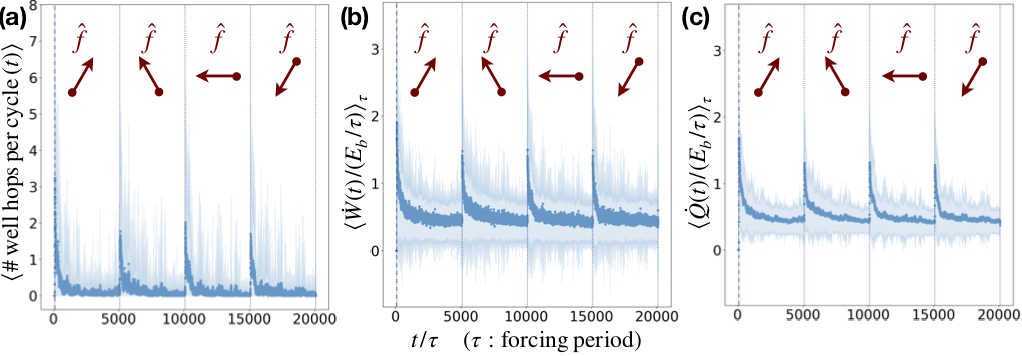}
\caption{Changing forcing direction by $\pi/3$. Each rotation of the forcing direction leads to a spike in the number of springs that change configuration shown in (a), because of a spike in the work absorption rate shown in (b) and dissipation rate shown in (c). Eventually a stable configuration with low work absorption rate, and low dissipation rate is attained. Averages are taken over 120 networks.}
\label{switching_dirn}
\end{figure*}

\section{Contraction analysis}
We use the results of contraction analysis \cite{lohmiller_contraction_1998} to infer that the low energy motion of a stable configuration of the mechanical network, i.e. motion within a potential energy well, is converging to a single one-dimensional trajectory in phase space, in sharp contrast to thermalized motion which corresponds to a non-vanishing volume in phase space.

On account of the low temperature of the surrounding thermal bath, the thermal noise term in the equation of motion can be neglected, making the system dynamics deterministic like that of a nonlinear Hamiltonian system subject to weak damping and external driving. A tool for studying driven or time-varying nonlinear dynamical systems is contraction analysis ~\cite{lohmiller_contraction_1998}, which asks whether a region of phase space is a contraction region, i.e., a region where all trajectories in the region converge towards one another and hence to a single trajectory.
If the system is driven by a periodic input, then all trajectories within
the contraction region converge to a single periodic trajectory with the same time-period as the input~\cite{lohmiller_contraction_1998}, and therefore to a one-dimensional closed trajectory in phase space.

Neglecting thermal noise, we can rewrite the equation of the motion in terms of the positions and momenta $\{ \vec{q}_i\,,\, \vec{p}_i \}$ of the particles as follows:
\begin{align}
    & \dot{\vec{q}}_i = \frac{\vec{p}_i}{m} \;,\; \dot{\vec{p}}_i = - \frac{\partial \mathcal{V}}{\partial \vec{q}_i} - \frac{\gamma}{m}\vec{p}_i + \vec{F}_i(t) \label{phase_space_evolution_no_noise} \\
    & \mathcal{V} = \sum_{i=1}^{N}\sum_{j=i+1}^{N}A_{ij}\,U(\vert \vec{x}_i -\vec{x}_j \vert) \nonumber
\end{align}
where the damping rate $\gamma$ and mass $m$ are strictly positive constants. Every stable equilibrium in the phase space corresponds to the potential energy landscape $\mathcal{V}(q)$ being locally concave at that point. For each stable equilibrium, one can find a constant uniformly positive definite metric $\bf{M}$ such that the Jacobian matrix $\bf{J}$ of the linearized system at this point verifies
the Lyapunov matrix equation $ \ \bf{M \ J} < \bf{0} $.
Using this same constant metric $\bf{M}$, this implies in turn that there is an entire finite neighborhood of the equilibrium such that 
$ \ \bf{M \ J}(\bf {x} ) < \bf{0} $, 
where $\bf {x}$ denotes the system state and $\bf{J}(\bf {x})$ the Jacobian matrix computed at that state. This neighborhood is therefore
a contraction region, and any two trajectories which remain within it tend exponentially to one another~\cite{lohmiller_contraction_1998}.

Since at long times, the system is typically trapped in a metastable configuration i.e. in a concave-up region of the potential energy landscape, all trajectories in the neighborhood of the system trajectory at long times are converging towards a single trajectory determined by its potential well and external drive, that is periodic with the same period as the driving period (see \cite{lohmiller_contraction_1998} Sec. 3.7(vi)). We studied the stability of this particular trajectory compared to other particular trajectories corresponding to different external drives within the same potential energy well, in Sec. 3B of the paper.

\bibliographystyle{unsrt}
\bibliography{refs}

\begin{thebibliography}{10}

\bibitem{tsamados_plasticity_2010}
M.~Tsamados.
\newblock Plasticity and dynamical heterogeneity in driven glassy materials.
\newblock {\em The European Physical Journal E}, 32(2):165--181, June 2010.

\bibitem{henderson_metastability_1996}
W.~Henderson, E.~Y. Andrei, M.~J. Higgins, and S.~Bhattacharya.
\newblock Metastability and {Glassy} {Behavior} of a {Driven} {Flux}-{Line}
  {Lattice}.
\newblock {\em Physical Review Letters}, 77(10):2077--2080, September 1996.

\bibitem{berthier_non-equilibrium_2013}
Ludovic Berthier and Jorge Kurchan.
\newblock Non-equilibrium glass transitions in driven and active matter.
\newblock {\em Nature Physics}, 9(5):310--314, May 2013.

\bibitem{keim_generic_2011}
Nathan~C. Keim and Sidney~R. Nagel.
\newblock Generic {Transient} {Memory} {Formation} in {Disordered} {Systems}
  with {Noise}.
\newblock {\em Physical Review Letters}, 107(1):010603, June 2011.

\bibitem{keim_multiple_2013}
Nathan~C. Keim, Joseph~D. Paulsen, and Sidney~R. Nagel.
\newblock Multiple transient memories in sheared suspensions: {Robustness},
  structure, and routes to plasticity.
\newblock {\em Physical Review E}, 88(3):032306, September 2013.

\bibitem{paulsen_multiple_2014}
Joseph~D. Paulsen, Nathan~C. Keim, and Sidney~R. Nagel.
\newblock Multiple {Transient} {Memories} in {Experiments} on {Sheared}
  {Non}-{Brownian} {Suspensions}.
\newblock {\em Physical Review Letters}, 113(6):068301, August 2014.

\bibitem{harne_designing_2015}
R.~L. Harne, Z.~Wu, and K.~W. Wang.
\newblock Designing and {Harnessing} the {Metastable} {States} of a {Modular}
  {Metastructure} for {Programmable} {Mechanical} {Properties} {Adaptation}.
\newblock {\em Journal of Mechanical Design}, 138(2):021402--021402--9,
  December 2015.

\bibitem{hertz_introduction_1991}
John Hertz, Anders Krogh, and Richard~G. Palmer.
\newblock {\em Introduction to the {Theory} of {Neural} {Computation}}.
\newblock Addison-Wesley Longman Publishing Co., Inc., Boston, MA, USA, 1991.

\bibitem{stern_dynamics_2014}
M.~Stern, H.~Sompolinsky, and L.~F. Abbott.
\newblock Dynamics of random neural networks with bistable units.
\newblock {\em Physical Review E}, 90(6):062710, December 2014.

\bibitem{cheng_multistability_2006}
Chang-Yuan Cheng, Kuang-Hui Lin, and Chih-Wen Shih.
\newblock Multistability in {Recurrent} {Neural} {Networks}.
\newblock {\em SIAM Journal on Applied Mathematics}, 66(4):1301--1320, 2006.

\bibitem{bacot_multistable_2019}
V.~Bacot, S.~Perrard, M.~Labousse, Y.~Couder, and E.~Fort.
\newblock Multistable {Free} {States} of an {Active} {Particle} from a
  {Coherent} {Memory} {Dynamics}.
\newblock {\em Physical Review Letters}, 122(10):104303, March 2019.

\bibitem{pisarchik_experimental_2003}
A.~N. Pisarchik, Yu.~O. Barmenkov, and A.~V. Kir’yanov.
\newblock Experimental demonstration of attractor annihilation in a multistable
  fiber laser.
\newblock {\em Physical Review E}, 68(6):066211, December 2003.

\bibitem{pisarchik_control_2002}
A.~N. Pisarchik and B.~F. Kuntsevich.
\newblock Control of multistability in a directly modulated diode laser.
\newblock {\em IEEE Journal of Quantum Electronics}, 38(12):1594--1598,
  December 2002.

\bibitem{fern_energy_2016}
Joshua Fern, Jennifer Lu, and Rebecca Schulman.
\newblock The {Energy} {Landscape} for the {Self}-{Assembly} of a
  {Two}-{Dimensional} {DNA} {Origami} {Complex}.
\newblock {\em ACS Nano}, 10(2):1836--1844, February 2016.

\bibitem{ravelet_multistability_2004}
Florent Ravelet, Louis Marié, Arnaud Chiffaudel, and François Daviaud.
\newblock Multistability and {Memory} {Effect} in a {Highly} {Turbulent}
  {Flow}: {Experimental} {Evidence} for a {Global} {Bifurcation}.
\newblock {\em Physical Review Letters}, 93(16):164501, October 2004.

\bibitem{power_multiple_1993}
S.~B. Power and R.~Kleeman.
\newblock Multiple {Equilibria} in a {Global} {Ocean} {General} {Circulation}
  {Model}.
\newblock {\em Journal of Physical Oceanography}, 23(8):1670--1681, August
  1993.

\bibitem{rahmstorf_multiple_1995}
Stefan Rahmstorf.
\newblock Multiple {Convection} {Patterns} and {Thermohaline} {Flow} in an
  {Idealized} {OGCM}.
\newblock {\em Journal of Climate}, 8(12):3028--3039, December 1995.

\bibitem{braun_attractors_2010}
Jochen Braun and Maurizio Mattia.
\newblock Attractors and noise: {Twin} drivers of decisions and multistability.
\newblock {\em NeuroImage}, 52(3):740--751, September 2010.

\bibitem{marin_high_2013}
Bóris Marin, William~H. Barnett, Anca Doloc-Mihu, Ronald~L. Calabrese, and
  Gennady~S. Cymbalyuk.
\newblock High {Prevalence} of {Multistability} of {Rest} {States} and
  {Bursting} in a {Database} of a {Model} {Neuron}.
\newblock {\em PLOS Computational Biology}, 9(3):e1002930, March 2013.

\bibitem{goldberg_epigenetics:_2007}
Aaron~D. Goldberg, C.~David Allis, and Emily Bernstein.
\newblock Epigenetics: {A} {Landscape} {Takes} {Shape}.
\newblock {\em Cell}, 128(4):635--638, February 2007.

\bibitem{laurent_multistability:_1999}
Michel Laurent and Nicolas Kellershohn.
\newblock Multistability: a major means of differentiation and evolution in
  biological systems.
\newblock {\em Trends in Biochemical Sciences}, 24(11):418--422, November 1999.

\bibitem{wang_quantifying_2011}
Jin Wang, Kun Zhang, Li~Xu, and Erkang Wang.
\newblock Quantifying the {Waddington} landscape and biological paths for
  development and differentiation.
\newblock {\em Proceedings of the National Academy of Sciences},
  108(20):8257--8262, May 2011.

\bibitem{sittel_perspective:_2018}
Florian Sittel and Gerhard Stock.
\newblock Perspective: {Identification} of collective variables and metastable
  states of protein dynamics.
\newblock {\em The Journal of Chemical Physics}, 149(15):150901, October 2018.

\bibitem{knowlton_thresholds_1992}
Nancy Knowlton.
\newblock Thresholds and {Multiple} {Stable} {States} in {Coral} {Reef}
  {Community} {Dynamics}.
\newblock {\em Integrative and Comparative Biology}, 32(6):674--682, December
  1992.

\bibitem{scheffer_catastrophic_2001}
Marten Scheffer, Steve Carpenter, Jonathan~A. Foley, Carl Folke, and Brian
  Walker.
\newblock Catastrophic shifts in ecosystems.
\newblock {\em Nature}, 413(6856):591--596, October 2001.

\bibitem{feudel_complex_2008}
Ulrike Feudel.
\newblock Complex dynamics in multistable systems.
\newblock {\em International Journal of Bifurcation and Chaos},
  18(06):1607--1626, June 2008.

\bibitem{hopfield_neural_1982}
J.~J. Hopfield.
\newblock Neural networks and physical systems with emergent collective
  computational abilities.
\newblock {\em Proceedings of the National Academy of Sciences},
  79(8):2554--2558, April 1982.

\bibitem{canavier_nonlinear_1993}
C.~C. Canavier, D.~A. Baxter, J.~W. Clark, and J.~H. Byrne.
\newblock Nonlinear dynamics in a model neuron provide a novel mechanism for
  transient synaptic inputs to produce long-term alterations of postsynaptic
  activity.
\newblock {\em Journal of Neurophysiology}, 69(6):2252--2257, June 1993.

\bibitem{keim_memory_2019}
Nathan~C. Keim, Joseph~D. Paulsen, Zorana Zeravcic, Srikanth Sastry, and
  Sidney~R. Nagel.
\newblock Memory formation in matter.
\newblock {\em Reviews of Modern Physics}, 91(3):035002, July 2019.

\bibitem{kashiwagi_adaptive_2006}
Akiko Kashiwagi, Itaru Urabe, Kunihiko Kaneko, and Tetsuya Yomo.
\newblock Adaptive {Response} of a {Gene} {Network} to {Environmental}
  {Changes} by {Fitness}-{Induced} {Attractor} {Selection}.
\newblock {\em PLOS ONE}, 1(1):e49, December 2006.

\bibitem{bieling_force_2016}
Peter Bieling, Tai-De Li, Julian Weichsel, Ryan McGorty, Pamela Jreij,
  Bo~Huang, Daniel~A. Fletcher, and R.~Dyche Mullins.
\newblock Force {Feedback} {Controls} {Motor} {Activity} and {Mechanical}
  {Properties} of {Self}-{Assembling} {Branched} {Actin} {Networks}.
\newblock {\em Cell}, 164(1):115--127, January 2016.

\bibitem{majumdar_mechanical_2018}
Sayantan Majumdar, Louis C. Foucard, Alex J. Levine, and Margaret L. Gardel.
\newblock Mechanical hysteresis in actin networks.
\newblock {\em Soft Matter}, 14(11):2052--2058, 2018.

\bibitem{pashine_directed_2019}
Nidhi Pashine, Daniel Hexner, Andrea~J. Liu, and Sidney~R. Nagel.
\newblock Directed aging, memory and {Nature}'s greed.
\newblock {\em arXiv:1903.05776 [cond-mat]}, March 2019.
\newblock arXiv: 1903.05776.

\bibitem{england_dissipative_2015}
Jeremy~L. England.
\newblock Dissipative adaptation in driven self-assembly.
\newblock {\em Nature Nanotechnology}, 10(11):919--923, November 2015.

\bibitem{perunov_statistical_2016}
Nikolay Perunov, Robert~A. Marsland, and Jeremy~L. England.
\newblock Statistical {Physics} of {Adaptation}.
\newblock {\em Physical Review X}, 6(2):021036, June 2016.

\bibitem{yan_why_2013}
Le~Yan, Gustavo Düring, and Matthieu Wyart.
\newblock Why glass elasticity affects the thermodynamics and fragility of
  supercooled liquids.
\newblock {\em Proceedings of the National Academy of Sciences},
  110(16):6307--6312, April 2013.

\bibitem{pisarchik_annihilation_2000}
A.~N. Pisarchik and B.~K. Goswami.
\newblock Annihilation of {One} of the {Coexisting} {Attractors} in a
  {Bistable} {System}.
\newblock {\em Physical Review Letters}, 84(7):1423--1426, February 2000.

\bibitem{pisarchik_controlling_2001}
A.~N. Pisarchik.
\newblock Controlling the multistability of nonlinear systems with coexisting
  attractors.
\newblock {\em Physical Review E}, 64(4):046203, September 2001.

\bibitem{pisarchik_control_2014}
Alexander~N. Pisarchik and Ulrike Feudel.
\newblock Control of multistability.
\newblock {\em Physics Reports}, 540(4):167--218, July 2014.

\bibitem{goswami_control_2009}
B.~K. Goswami, S.~Euzzor, K.~Al~Naimee, A.~Geltrude, R.~Meucci, and F.~T.
  Arecchi.
\newblock Control of stochastic multistable systems: {Experimental}
  demonstration.
\newblock {\em Physical Review E}, 80(1):016211, July 2009.

\bibitem{goswami_controlled_2007}
B.~K. Goswami.
\newblock Controlled destruction of chaos in the multistable regime.
\newblock {\em Physical Review E}, 76(1):016219, July 2007.

\bibitem{gronbech-jensen_simple_2013}
Niels Gr{\o}nbech-Jensen and Oded Farago.
\newblock A simple and effective {Verlet}-type algorithm for simulating
  {Langevin} dynamics.
\newblock {\em Molecular Physics}, 111(8):983--991, April 2013.

\bibitem{prigogine_etude_1947}
I.~Prigogine.
\newblock {\em Etude thermodynamique des phénomènes irréversibles}.
\newblock PhD thesis, Brussels Univ, Bruxelles, 1947.

\bibitem{lohmiller_contraction_1998}
Winfried Lohmiller and Jean-Jacques~E. Slotine.
\newblock On {Contraction} {Analysis} for {Non}-linear {Systems}.
\newblock {\em Automatica}, 34(6):683--696, June 1998.

\bibitem{kondepudi_end-directed_2015}
Dilip Kondepudi, Bruce Kay, and James Dixon.
\newblock End-directed evolution and the emergence of energy-seeking behavior
  in a complex system.
\newblock {\em Physical Review E}, 91(5):050902(R), May 2015.

\bibitem{kachman_self-organized_2017}
Tal Kachman, Jeremy~A. Owen, and Jeremy~L. England.
\newblock Self-{Organized} {Resonance} during {Search} of a {Diverse}
  {Chemical} {Space}.
\newblock {\em Physical Review Letters}, 119(3):038001, July 2017.

\bibitem{chvykov_least-rattling_2018}
Pavel Chvykov and Jeremy England.
\newblock Least-rattling feedback from strong time-scale separation.
\newblock {\em Physical Review E}, 97(3):032115, March 2018.

\bibitem{volker_hole-burning_1989}
Silvia V{\"o}lker.
\newblock Hole-{Burning} {Spectroscopy}.
\newblock {\em Annual Review of Physical Chemistry}, 40(1):499--530, 1989.

\end{thebibliography}
\end{document}